\documentclass[twocolumn,preprintnumbers,amsmath,amssymb,prl,reprint,groupedaddress,
amsmath,citeautoscript,flushbottom,floatfix,superscriptaddress]{revtex4}
\usepackage{graphicx}
\usepackage{dcolumn}
\usepackage{bm}
\usepackage{amsmath}
\usepackage{amssymb}
\usepackage{booktabs}
\usepackage{multirow}
\usepackage{array}
\usepackage{color}
\usepackage{xcolor}
\usepackage[normalem]{ulem}
\usepackage{float}
\usepackage{booktabs}
\usepackage{diagbox}

\newcommand{\vc}[1]{\boldsymbol{#1}}

\begin{document}

\title{ Exchange interactions in $d^5$ Kitaev materials: From Na$_2$IrO$_3$ to $\alpha$-RuCl$_3$ }

\author{Huimei Liu}
\affiliation{Institute for Theoretical Solid State Physics and W$\ddot{u}$rzburg-Dresden Cluster of Excellence ct.qmat, IFW Dresden, Helmholtzstr. 20, 01069 Dresden, Germany}
\affiliation{Max Planck Institute for Solid State Research, Heisenbergstrasse 1, D-70569 Stuttgart, Germany}

\author{Ji\v{r}\'{\i} Chaloupka}
\affiliation{Department of Condensed Matter Physics, Faculty of Science, Masaryk University, Kotl\'a\v{r}sk\'a 2, 61137 Brno, Czech Republic}

\author{Giniyat Khaliullin}
\affiliation{Max Planck Institute for Solid State Research, Heisenbergstrasse 1, D-70569 Stuttgart, Germany}

\begin{abstract}
We present an analytical study of the exchange interactions between pseudospin one-half $d^5$ ions in honeycomb lattices with edge-shared octahedra. Various exchange channels involving Hubbard $U$, charge-transfer excitations, and cyclic exchange are considered. Hoppings within $t_{2g}$ orbitals as well as between $t_{2g}$ and $e_g$ orbitals are included. Special attention is paid to the trigonal crystal field $\Delta$ effects on the exchange parameters. The obtained exchange Hamiltonian is dominated by ferromagnetic Kitaev interaction $K$ within a wide range of $\Delta$. It is found that a parameter region close to the charge-transfer insulator regime and with a small $\Delta$ is most promising to realize the Kitaev spin liquid phase. Two representative honeycomb materials  Na$_2$IrO$_3$ and $\alpha$-RuCl$_3$ are discussed based on our theory. We have found that both materials share dominant ferromagnetic $K$ and positive non-diagonal $\Gamma$ values. However, their Heisenberg $J$ terms have opposite signs: AFM $J>0$ in  Na$_2$IrO$_3$ and FM $J<0$ in $\alpha$-RuCl$_3$. This brings different magnetic fluctuations and results in their different magnetization behaviors and spin excitation spectra. Proximity to FM state due to the large FM $J$ is emphasized in $\alpha$-RuCl$_3$. The differences between the exchange couplings of these two materials originate from the opposite $\Delta$ values, indicating that the crystal field can serve as an efficient control parameter to tune the magnetic properties of $d^5$ spin-orbit Mott insulators.
\end{abstract}

\date{\today}
\maketitle

\section{I. Introduction}

The exactly solvable Kitaev honeycomb model~\cite{Kit06} and its extensions have attracted much attention in recent years (see Refs. \cite{Rau16,Her18,Tre17,Win17,Sav17,Tak19,Jan19,Mot20,Tom21} for review). In this model, the nearest-neighbor (NN) spins $S=1/2$ interact via a simple Ising-type coupling $S_i^\gamma S_j^\gamma$, with a bond-dependent Ising axis $\gamma$ which takes mutually orthogonal directions ($x,y,z$) on the three adjacent NN bonds of the honeycomb lattice, see Fig.~\ref{fig:1}(a). Due to the strong frustration, the spins form a highly entangled quantum many-body state, which supports fractional excitations described by Majorana fermions~\cite{Kit06}.

Tremendous efforts have been made to materialize the Kitaev spin liquid state. Physically, the Ising-type anisotropy as in the Kitaev model is a hallmark of unquenched orbital magnetism. Since the orbitals are spatially anisotropic and bond-directional, they naturally lead to the bond-dependent exchange anisotropy between orbital moments. The anisotropy can be inherited by total angular momentum through spin-orbit coupling (SOC)~\cite{Kha05}. Spin-orbit Mott insulators such as 5$d^5$ iridates~\cite{Jac09}, 4$d^5$ ruthenates~\cite{Plu14} and 3$d^7$ cobaltates~\cite{Liu18,San18} with pseudospin $\widetilde{S} =1/2$ ground state have been suggested to host the Kitaev model.

Strong bond-directional Kitaev interaction has indeed been reported in several materials such as Na$_2$IrO$_3$ \cite{Yam14,Chu15,Win16,Kim20} and $\alpha$-RuCl$_3$ (hereafter RuCl$_3$) \cite{Ban16,Win18,Win17b,Sea20,Lau20,Mak20,Suz21}. Instead of forming the Kitaev spin liquid state, however, these materials display long range magnetic orders at sufficiently low temperatures. This is driven by corrections to the Kitaev honeycomb model. A broad consensus on the form of minimal exchange Hamiltonian has been reached, namely the extended Kitaev model $\mathcal{H}_{ij}^{(\gamma)}$, which consists of symmetry allowed Kitaev $K$, Heisenberg $J$, and off-diagonal $\Gamma$ and $\Gamma'$ interactions between NN ions~\cite{Rau14a,Rau14b,Kat14}. Specifically, on the $z$-type NN bonds [see Fig.~\ref{fig:1}(a)], $\mathcal{H}_{ij}^{(z)}$ reads as
\begin{align}
\mathcal{H}_{ij}^{(z)} =& \; K \widetilde{S}_i^z\widetilde{S}_j^z +
J  \widetilde{\vc S}_i \cdot  \widetilde{\vc S}_j
+ \Gamma (\widetilde{S}_i^x\widetilde{S}_j^y +
\widetilde{S}_i^y\widetilde{S}_j^x )
\notag \\
&+ \Gamma'(\widetilde{S}_i^x\widetilde{S}_j^z +
\widetilde{S}_i^z\widetilde{S}_j^x+
\widetilde{S}_i^y\widetilde{S}_j^z +
\widetilde{S}_i^z\widetilde{S}_j^y) \;.
\label{eq:HK}
\end{align}
The interactions on $x$- and $y$-type NN bonds can be obtained by cyclic permutations among $\widetilde{S}^x$, $\widetilde{S}^y$, and $\widetilde{S}^z$. In addition to Eq.~\eqref{eq:HK}, the longer-range spin interactions are present in real materials. Albeit much weaker than NN Kitaev coupling $K$, they are often included for a quantitative description of the experimental data.

While the overall structure of spin Hamiltonian in Kitaev materials is fixed by underlying lattice symmetry and thus rather generic, the specific values of coupling constants $K$, $J$, $\Gamma$, and $\Gamma'$ are sensitive to material's chemistry and may vary broadly. This results in a diversity of magnetic properties: various magnetic orderings and excitation spectra, different responses to external magnetic field, etc. \cite{Tak19,Jan19}. Particularly, the competition between the Kitaev and non-Kitaev terms decides the proximity of a given compound to the Kitaev spin liquid phase. Therefore it is important to develop a quantitative theory of the exchange interactions in Kitaev materials, and
understand how the  ``undesired'' non-Kitaev couplings depend on the material intrinsic properties, such as the interplay between different hopping channels, strength and sign of non cubic crystal fields, and so on.

In this paper, we present a systematic microscopic derivation of the exchange interaction parameters for honeycomb $d^5$ spin-orbit Mott insulators. We consider hopping channels not only within $t_{2g}$ orbitals, but also involving $e_g$ orbitals. The inclusion of $e_g$ orbitals into the exchange processes is important for the quantitative values of $K$ and $J$. According to our calculations, the trigonal crystal fields, present in real materials, have a particularly large effect on the exchange interactions, and small $\Delta$ materials are favored to host dominant Kitaev interaction. We also calculated the exchange parameters as a function of the ratio between Hubbard $U$ and charge-transfer gap $\Delta_{pd}$, and found that to realize the Kitaev spin liquid phase, a parameter regime close to charge-transfer limit ($U \geq \Delta_{pd}$) is desirable.

As the test cases, we have applied our theory to two representative Kitaev materials, Na$_2$IrO$_3$ and RuCl$_3$. We found that both materials have dominant ferromagnetic (FM) $K$  and sizable positive $\Gamma$ interactions. However, they are characterized by opposite signs of the Heisenberg coupling $J$ (AFM $J>0$ in  Na$_2$IrO$_3$ and FM $J<0$ in RuCl$_3$), which is responsible for different  magnetization behaviors and magnetic excitation spectra. The opposite signs of $J$ values can be traced back to the opposite signs of trigonal crystal field in Na$_2$IrO$_3$ and RuCl$_3$. This indicates that trigonal crystal field could serve as an efficient tuning parameter of the exchange interactions in $d^5$ materials, as in the case of $3d^7$ cobaltates~\cite{Liu20,Liu21}.

The paper is organized as follows. Section II presents the detailed derivations of the general exchange Hamiltonian. The microscopic origins of the coupling constants are systematically studied. A parameter regime with the possibility of realizing the Kitaev spin liquid phase is identified. Section III presents the application of our theory to two materials: Na$_2$IrO$_3$ and RuCl$_3$. The exchange parameters are obtained for both materials. The spin excitation spectra, calculated by linear spin-wave theory and the exact diagonalization method, are compared with experimental data. The paper is summarized in Sec. IV.

\section {II. Exchange interactions between Pseudospins $\widetilde{S}=1/2$
under trigonal crystal field}
\begin{figure}
\begin{center}
\includegraphics[width=8.5cm]{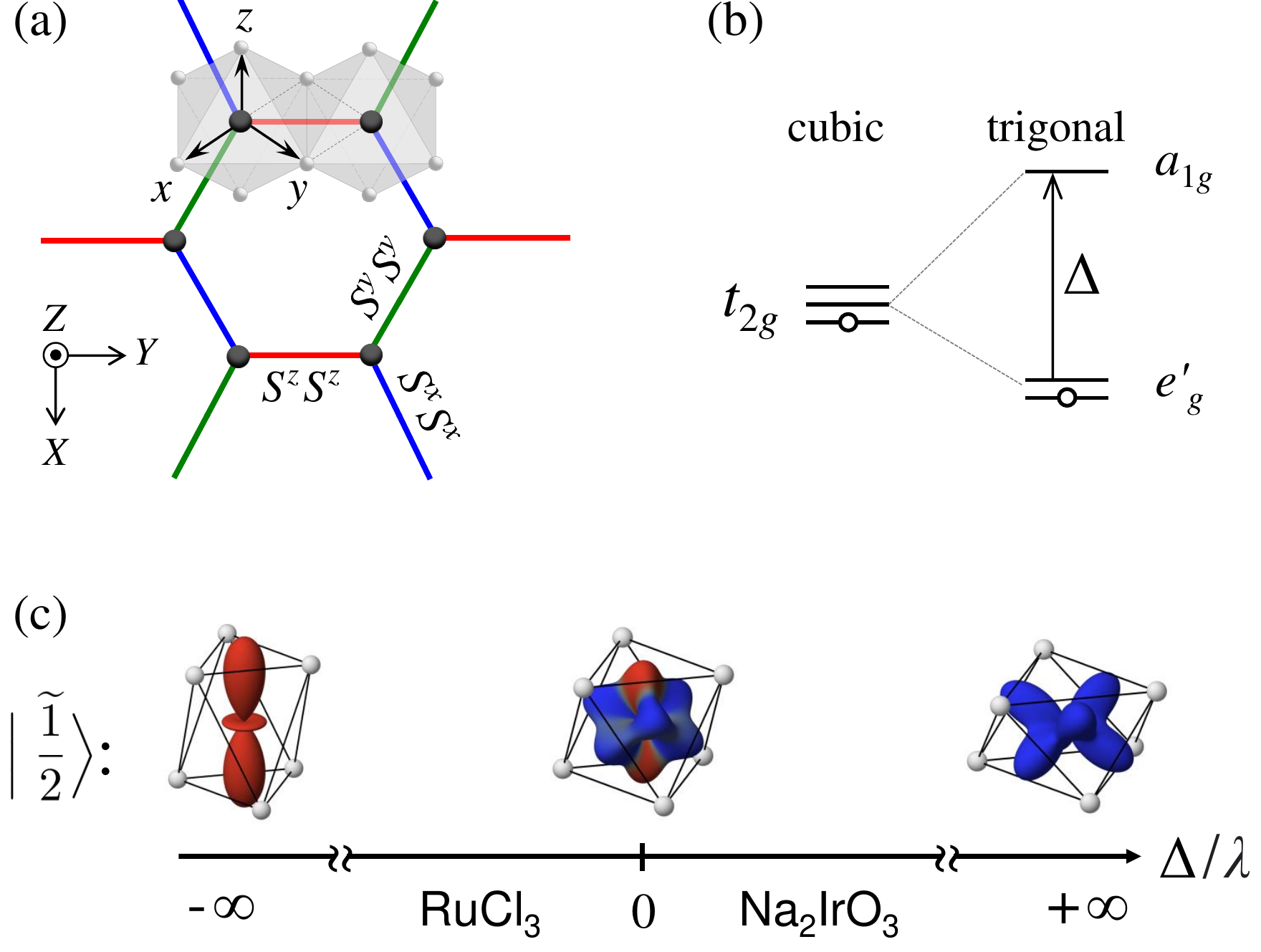}
\caption{(a) Top view of the honeycomb plane. NN bonds hosting  $x$-, $y$- and $z$-type Ising couplings are shown in
blue, green, and red colors, respectively. The hexagonal lattice coordinates ($X, Y, Z$) and octahedral axes ($x, y, z$) are indicated.
(b) Level structure of $t_{2g}$ manifold in a hole representation.
Positive $\Delta>0$ corresponds to a compression of octahedra along $Z$-axis within a point-charge model.
(c) Wavefunction shapes of pseudospin $\Big|  \widetilde{\frac{1}{2}}\Big\rangle$  state in cubic case ($\Delta=0$) and two opposite limits of trigonal crystal field $\Delta$.
Red (blue) color indicates the spin up (down) polarization of the hole. Na$_2$IrO$_3$ and RuCl$_3$ have opposite signs of $\Delta$, thus the shapes of their ground state wavefunctions should be different and result in the distinct exchange interactions and magnetic spectra (see text).}
\label{fig:1}
\end{center}
\end{figure}

In Na$_2$IrO$_3$ and RuCl$_3$, the transition metal ions Ir$^{4+}$ and Ru$^{3+}$ both possess a $d^5$ electronic configuration with five electrons residing on $t_{2g}$ orbitals, forming $S=1/2$ and an effective $L=1$ orbital moments. The trigonal crystal field $\Delta$ splits the $t_{2g}$ orbitals into a singlet
$a_{1g}$ corresponding to the $L_Z=0$ state, and a doublet $e'_g$ hosting the $L_Z=\pm 1$ states, see Fig.~\ref{fig:1}(b). In terms of the effective angular momentum $L=1$ of the $t^5_{2g}$ configuration \cite{Abr70},
the relations between the $|L_z\rangle$ states and orbitals hold as
$|0\rangle  =\tfrac{1}{\sqrt{3}} \left(|a\rangle+|b\rangle+|c\rangle \right)$
and $| \pm 1\rangle  = \pm \tfrac{1}{\sqrt{3}}  (e^{\pm i\tfrac{2\pi}{3}}|a\rangle
+e^{\mp i\tfrac{2\pi}{3}}|b\rangle+|c\rangle  )$, where
the shorthand notations $a= d_{yz}$,  $b= d_{zx}$ and  $c= d_{xy}$ are introduced.

Under SOC $H_{\lambda}=\lambda \vc L \cdot \vc S$ and trigonal crystal field $H_{\Delta}= \Delta(L_Z^2-2/3)$, the ground state Kramers doublet hosts the pseudospin $\widetilde{S} =1/2$ state, with the wavefunctions written in the $|L_Z,S_Z \rangle$ basis
as follows:
\begin{align}
\Big|\!\pm \widetilde{\frac{1}{2}}\Big\rangle& =
\pm s_{\theta} \Big|0,\pm\frac{1}{2} \Big\rangle
\mp c_{\theta} \Big|\pm 1,\mp\frac{1}{2} \Big\rangle.
\label{eq:wf}
\end{align}
The coefficients $s_{\theta} \equiv \sin \theta$ and $c_{\theta} \equiv \cos \theta$, and the spin-orbit mixing angle $\theta$ is determined by
$\tan 2\theta=2\sqrt{2}/(1+\delta)$ with $\delta=2\Delta/\lambda$. In the cubic limit ($\Delta=0$), we have $s_{\theta}=1/\sqrt{3}$ and $c_{\theta}=\sqrt{2/3}$,
and the three $t_{2g}$ orbitals contribute equally to the wave functions.

The trigonal field modifies the shape of the ground state wave functions as shown in Fig.~\ref{fig:1}(c). We will see that this modification strongly affects the exchange parameters. To obtain the exchange parameters between pseudospins-1/2, we first need to derive the Kugel-Khomskii type spin-orbital exchange Hamiltonian, and then project it onto the ground state doublet subspace defined by wave functions Eq.~(\ref{eq:wf}).

We divide the exchange processes into two classes: exchange between (1) $t_{2g}$ and $t_{2g}$ orbitals, and (2) $t_{2g}$ and $e_g$ orbitals, both channels being relevant for the 90$^{\circ}$ bonding geometry of the edge-shared octahedra. Within each class, three exchange mechanisms are considered. As illustrated in Fig.~\ref{fig:2}, they involve \\
\textbf{(i)} Mott-Hubbard transitions with excitation energy $U$,
\vskip 2mm
\noindent \textbf{(ii)} charge-transfer excitations with energy $\Delta_{pd}$, and
\vskip 2mm
\noindent \textbf{(iii)} cyclic-exchange mechanism.

Since the ground state wave functions Eq.~(\ref{eq:wf}) are defined in the hexagonal $XYZ$ basis, it is technically easier to obtain the pseudospin exchange Hamiltonian also in the $XYZ$ coordinate frame defined in Fig.~\ref{fig:1}(a). By symmetry, the exchange Hamiltonian between pseudospins $\widetilde{S} =1/2$ has the following general form \cite{Cha15}:
\begin{align}
\mathcal{H}_{ij}^{(\gamma)}= & J_{XY}
\left(\widetilde{S}_i^X\widetilde{S}_j^X +
\widetilde{S}_i^Y\widetilde{S}_j^Y \right) +
J_Z \widetilde{S}_i^Z\widetilde{S}_j^Z  \notag \\
+ & A \left[ c_{\gamma}\left(\widetilde{S}_i^X\widetilde{S}_j^X  \!-\!
\widetilde{S}_i^Y\widetilde{S}_j^Y \right)\!-\!
s_{\gamma}\left(\widetilde{S}_i^X\widetilde{S}_j^Y  \!+\!
\widetilde{S}_i^Y\widetilde{S}_j^X \right) \right]   \notag \\
- & B \sqrt {2}\left[ c_{\gamma}
\left(\widetilde{S}_i^X\widetilde{S}_j^Z \!+\!
\widetilde{S}_i^Z\widetilde{S}_j^X \right)
\!+\! s_{\gamma}\left(\widetilde{S}_i^Y\widetilde{S}_j^Z \!+\!
\widetilde{S}_i^Z\widetilde{S}_j^Y \right) \right],
\label{eq:H}
\end{align}
with $c_{\gamma}\equiv \cos \phi_{\gamma}$ and $s_{\gamma}\equiv \sin \phi_{\gamma}$. The angles $\phi_{\gamma}=(0, 2 \pi /3 , 4 \pi /3)$ refer to the $\gamma=z$-, $x$-, and $y$-type NN bonds in Fig.~\ref{fig:1}(a), respectively.

\textcolor[rgb]{0.00,0.00,0.00}{One can convert Eq.~(\ref{eq:H}) into the more familiar form, namely, the extended Kitaev model of Eq.~(\ref{eq:HK}), written in the octahedral $xyz$ coordinate frame. The corresponding} exchange parameters $K$, $J$, $\Gamma$, and $\Gamma'$ are related to $J_{XY}$, $J_Z$, $A$, and $B$ of Eq.~(\ref{eq:H}) as follows:
\begin{alignat}{2}
K &=A+2B \; ,
\notag \\
J &=\tfrac{1}{3}(2J_{XY}+J_Z)-\tfrac{1}{3}(A+2B) \; ,
\notag \\
\Gamma &=\tfrac{1}{3}(J_Z-J_{XY})+\tfrac{2}{3}(A-B)  \; ,
\notag \\
\Gamma' &=\tfrac{1}{3}(J_Z-J_{XY})-\tfrac{1}{3}(A-B) \; .
\label{eq:Tr}
\end{alignat}
In the following text, we will skip the intermediate calculation steps
and show the exchange parameters directly in the form as defined in Eq.~(\ref{eq:HK}).

\subsection{2.1. Exchange between $t_{2g}$ and $t_{2g}$ orbitals}

\begin{figure}
\begin{center}
\includegraphics[width=8.5cm]{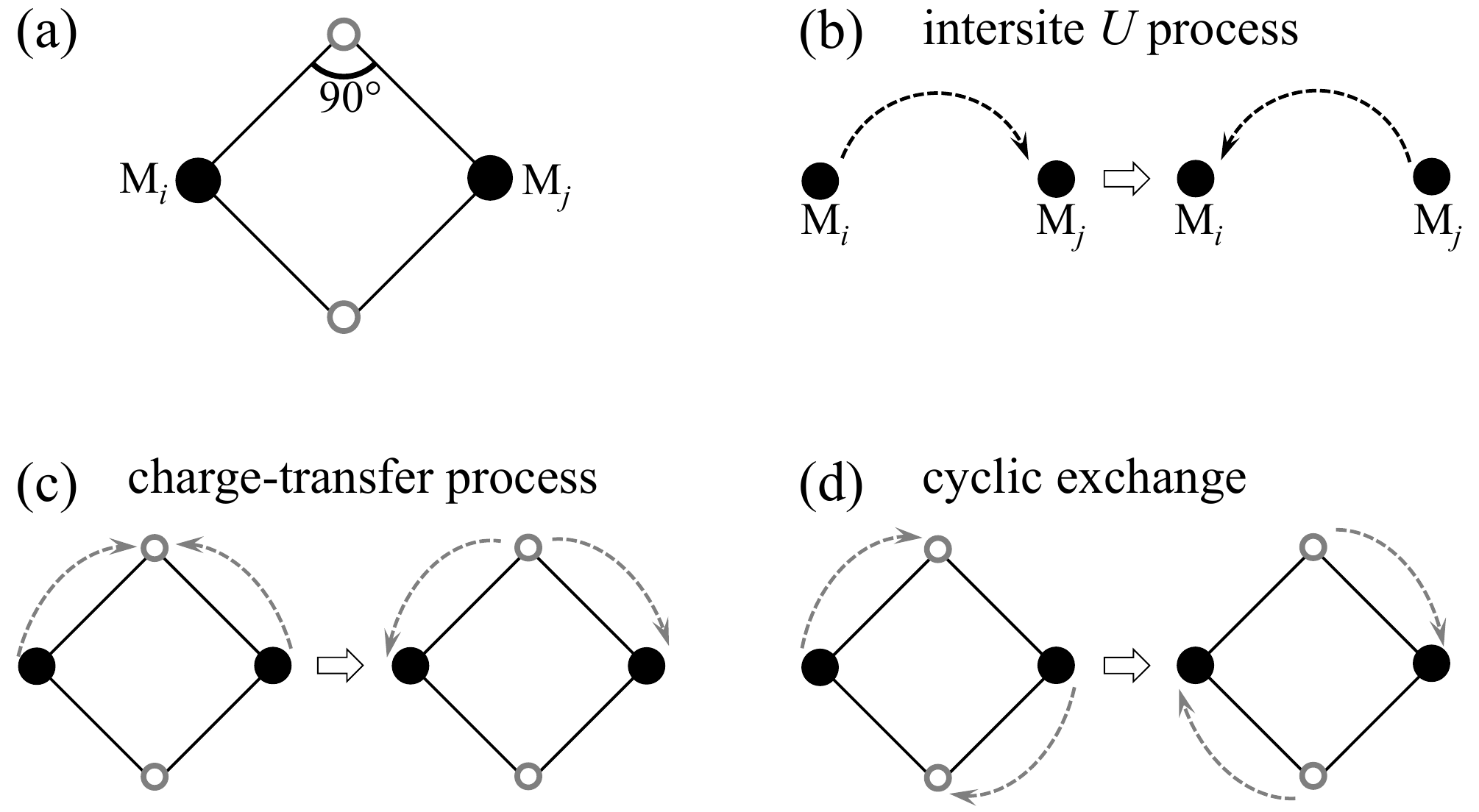}
\caption{(a) The 90$^{\circ}$ bonding geometry between the neighboring
magnetic ions M$_i$ and M$_j$. Sketches of the considered (b) intersite $U$ process where two holes meet at the transition metal ion site,
(c) charge-transfer process where two holes are created at the same ligand ion, and (d) cyclic exchange where two holes are created on different ligand ions and do not meet each other.}
\label{fig:2}
\end{center}
\end{figure}

\begin{figure}
\begin{center}
\includegraphics[width=8.5cm]{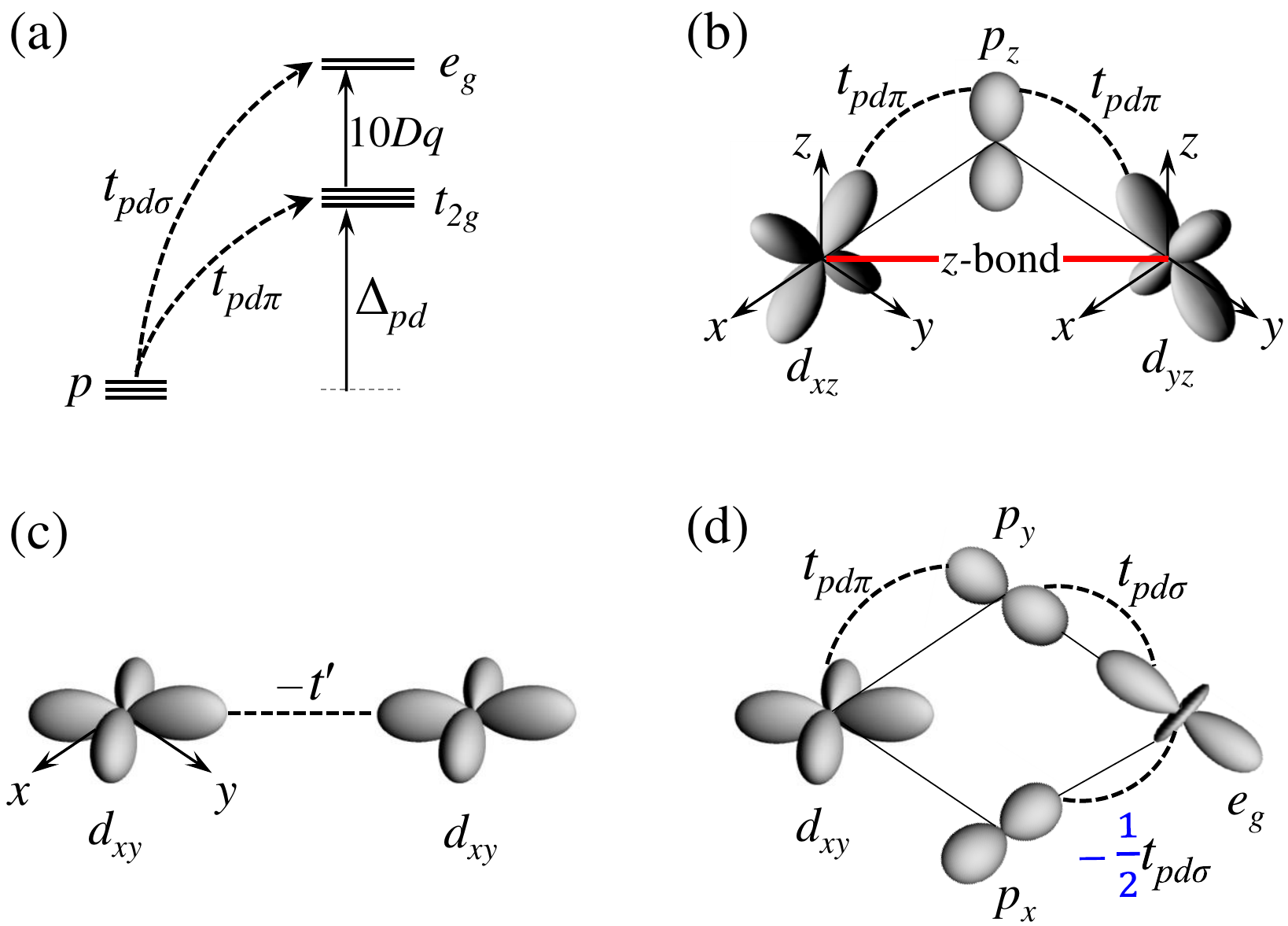}
\caption{ (a) Schematic of $p$ and transition-metal ion
$d$($t_{2g}$,$e_g$) energy levels. $t_{pd\pi}$($t_{pd\sigma}$) is the hopping integral between $p$ and $t_{2g}$($e_g$) orbitals. The corresponding $pd$ charge-transfer gap $\Delta_{pd}$ and cubic splitting $10Dq$ are indicated.
Sketch of different hopping processes between $d$ orbitals along $z$-type NN-bond:
(b) indirect hopping $t$ between $t_{2g}$ orbitals,
(c) direct hopping $t'$ between $c=d_{xy}$ orbitals, and
(d) indirect hopping processes between $t_{2g}$ and $e_g$ orbitals. Notice that there is a prefactor ($-1/2$) of the overlap between $e_g$ orbital and lower ligand $p_x$ orbital in (d).
}
\label{fig:3}
\end{center}
\end{figure}

In a 90$^{\circ}$ bonding geometry as shown in Fig.~\ref{fig:2}(a), the hopping between  $t_{2g}$ orbitals along the
$\gamma=z$-type NN-bonds can be written as~\cite{Kha04,Kha05,Nor08,Cha11}:
\begin{align}
\mathcal{H}^{(z)}_t = \sum_{\sigma} \left[t(a_{i\sigma}^{\dag}b_{j\sigma}+b_{i\sigma}^{\dag}a_{j\sigma})
-t'c_{i\sigma}^{\dag}c_{j\sigma} + \mathrm{H.c} \right].
\label{eq:tt'}
\end{align}
Here $\sigma$ is spin index, $t= t_{pd\pi}^2/\Delta_{pd}$ is the indirect hopping between $a= d_{yz}$ and $b= d_{zx}$ orbitals through the ligand $p$-states via the $p$-$d$ charge-transfer gap $\Delta_{pd}$.
 $t'>0$ is the direct overlap between $c= d_{xy}$ orbitals, see Figs.~\ref{fig:3}(a)-\ref{fig:3}(c).

\subsubsection{2.1.1 Intersite $U$ processes }
In the intersite $U$ process, virtual charge transitions are of the $d^5_i d^5_j \rightarrow d^4_i d^6_j$ type, i.e. transitions across the Mott-Hubbard gap are involved, as shown in Fig.~\ref{fig:2}(b).
The corresponding spin-orbital exchange Hamiltonian for $\gamma=z$-type NN-bonds is given by
\begin{align}
\mathcal{H}^{(z)}_{11} =&-t^2
\left(\frac{P_T}{E_1}+\frac{P_S}{E_2} \right)(n_{ia}+n_{ib}+n_{ja}+n_{jb} )
\notag \\
& +2t^2\left(\frac{P_T}{E_1}-\frac{P_S}{E_2} \right) (n_{ia}n_{jb}+a_i^{\dag}b_ia_j^{\dag}b_j+a\leftrightarrow b)
\notag \\
& +\frac{4}{3}t^2\left(\frac{1}{E_2}-\frac{1}{E_3} \right)P_S (n_{ia}n_{jb}+a_i^{\dag}b_ib_j^{\dag}a_j+a\leftrightarrow b)
\notag \\
&-2tt'\left(\frac{P_T}{E_1}-\frac{P_S}{E_2} \right) (a_i^{\dag}c_ic_j^{\dag}b_j +c_i^{\dag}a_ib_j^{\dag}c_j+a\leftrightarrow b)
\notag \\
&-\frac{2}{3}tt'\left(\frac{1}{E_2}-\frac{1}{E_3} \right)P_S  (a_i^{\dag}c_ib_j^{\dag}c_j
+c_i^{\dag}a_ic_j^{\dag}b_j+a\leftrightarrow b)
\notag \\
&-\frac{4}{3}t'^2\left(\frac{2}{E_2}+\frac{1}{E_3} \right)P_S n_{ic}n_{jc}
\notag \\
&-t'^2\left(\frac{P_T}{E_1}+\frac{P_S}{E_2} \right)(n_{ic}+n_{jc}-2n_{ic}n_{jc})
\;.
\label{eq:H1}
\end{align}
Here $n_a=a^{\dag}a$, etc. are the orbital occupations,
$P_T=\tfrac{3}{4}+(\vc S_i \cdot \vc S_j)$ and
$P_S=\tfrac{1}{4}-(\vc S_i \cdot \vc S_j)$
are spin triplet and singlet state projectors, respectively.
The excitation energies are represented by
a high-spin transition at $E_1=U-3J_H$ and low-spin transitions at  $E_2=U-J_H$ and $E_3=U+2J_H$,
where $U$ and $J_H$ are the Coulomb interaction and Hund's coupling on $d$ orbitals.

Next step is to project Eq.~(\ref{eq:H1}) onto pseudospin $\widetilde{S}=1/2$ subspace. To this end, we calculate the matrix elements of spin-orbital operators within the pseudospin
$\widetilde{S}=1/2$ doublet Eq. (\ref{eq:wf}), and obtain the following operator correspondences:
\begin{align}
S_{X/Y} = -s^2_{\theta} \widetilde{S}_{X/Y} \;, \ \ \ \ \
S_Z = -c_{2\theta}\widetilde{S}_Z \;.
\label{eq:tai}
\end{align}
\begin{align}
a^{\dagger}b  & = \tfrac{i}{\sqrt{3}} \left( c^2_{\theta} \widetilde{S}_Z
- s_{ 2\theta} \widetilde{S}_X \right)  \;,
\notag \\
 (a^{\dagger}/b^{\dagger})c &   =\mp \tfrac{i}{\sqrt{3}}  \left( c^2_{\theta}  \widetilde{S}_Z +  s_{s\theta}  \widetilde{S}_{A/B} \right)   \;,
\end{align}
\textcolor[rgb]{0.00,0.00,0.00}{where $\widetilde{S}_A = \tfrac{1}{2}( \widetilde{S}_X+ \sqrt{3}\widetilde{S}_Y )$ and $\widetilde{S}_B = \tfrac{1}{2}( \widetilde{S}_X- \sqrt{3}\widetilde{S}_Y )$.}
\begin{align}
  S_X n_{a/b} &   =\tfrac{1}{3} \left( c^2_{\theta} \widetilde{S}_{B/A}
+\tfrac{1}{2} s_{2\theta}\widetilde{S}_Z \right) - \tfrac{1}{3}  \widetilde{S}_X   \;,
\notag \\
  S_Y n_{a/b} &    =\tfrac{1}{\sqrt{3}}  \left( \mp  c^2_{\theta}\widetilde{S}_{B/A}
+ \tfrac{1}{2} s_{2\theta}\widetilde{S}_Z \right) - \tfrac{1}{3}  \widetilde{S}_Y   \;,
\notag \\
  S_Z n_{a/b} &   =\tfrac{1}{3} \left(  s_{2\theta}\widetilde{S}_{B/A}
-c_{2\theta}\widetilde{S}_Z  \right)  \;, \notag \\
S_{X/Z} n_c & =\pm \tfrac{1}{3} \left(c_{2\theta} \widetilde{S}_{X/Z} \mp s_{2\theta} \widetilde{S}_{Z/X}\right)\;,
\notag \\
S_Y n_c &=-\tfrac{1}{3} \widetilde{S}_Y\;,
\end{align}
and
\begin{align}
S_X a^{\dagger}b & =\tfrac{1}{3} c_{2\theta} \widetilde{S}_X+\tfrac{1}{6} s_{2\theta} \widetilde{S}_Z \;,
\notag \\
S_Y a^{\dagger}b  &=-\tfrac{1}{3} \widetilde{S}_Y  \;,
\notag \\
S_Z a^{\dagger}b &=- \tfrac{1}{6} \left( c^2_{\theta} \widetilde{S}_Z
- s_{ 2\theta} \widetilde{S}_X \right) + \tfrac{1}{3} \widetilde{S}_Z   \;,
\end{align}
and
\begin{align}
 S_X (a^{\dagger}/b^{\dagger})c &   =\tfrac{1}{3} \left( c^2_{\theta} \widetilde{S}_{A/B}
-\tfrac{1}{4} s_{2\theta}\widetilde{S}_Z \right) - \tfrac{1}{3}  \widetilde{S}_X     \;,
\notag \\
 S_Y (a^{\dagger}/b^{\dagger})c &    =\pm \tfrac{1}{\sqrt{3}}  \left(  c^2_{\theta}\widetilde{S}_{A/B}
- \tfrac{1}{4} s_{2\theta}\widetilde{S}_Z \right) - \tfrac{1}{3}  \widetilde{S}_Y  \;,
\notag \\
  S_Z (a^{\dagger}/b^{\dagger})c &    =-\tfrac{1}{6} s_{2\theta}\widetilde{S}_{A/B}
+ \tfrac{1}{6} \left( 1+s^2_{\theta} \right) \widetilde{S}_Z \;.
\label{eq:taf}
\end{align}

Using the projection table listed in Eqs.~(\ref{eq:tai})-(\ref{eq:taf}), one can convert Eq. (\ref{eq:H1}) into the form of Eq. (\ref{eq:HK}) with the exchange parameters:
\begin{flalign}
K_{11} \!=&\!-\frac{4}{3}\!\left(\!\frac{1}{E_1}\!-\!\frac{1}{E_2}\!\right)\!
\left[ (1+9 \alpha)(t^2\!-\!\tfrac{1}{3}t'^2)\!+\!
\mu_1 tt' \right],
\notag \\
J_{11} \!=& \frac{4}{27}\left(\!\frac{2}{E_2}\!+\!\frac{1}{E_3}\! \right)(t'+3 \beta t)^2
+\frac{4\beta}{9} \left(\!\frac{1}{E_2}\!-\!\frac{1}{E_3}\!\right) tt'
\notag \\
&\!+\!\frac{4}{3} \left(\!\frac{1}{E_1}\!-\!\frac{1}{E_2}\!\right)
[3\alpha t^2\!+\!(\mu_2\!+\!\beta )tt'
 \!-\! \mu_3 t'^2],
\notag \\
\Gamma_{11}  \!=& \frac{8}{9}  \left(\!\frac{1}{E_1}\!-\!\frac{1}{E_2}\!\right)
\left[(1+\mu_4) tt'+\mu_5 t^2
 -\tfrac{3}{4} \mu_2 t'^2 \right],
\notag \\
\Gamma'_{11}  \!=&\!-\frac{1}{3}  \left(\!\frac{1}{E_1}\!-\!\frac{1}{E_2}\! \right)
\left[ \mu_6 t^2+2\mu_7 tt' - \mu_7 t'^2)\right].
\label{eq:p1}
\end{flalign}
Here $\alpha=c^2_{\theta} \left(s^2_{\theta} + 1/\sqrt{2}  \right)/6-1/9 $, and
$\beta=c_{2\theta}/2-1/6$. Other parameters are: $\mu_1 = 6 \alpha\!+\! \beta\!+\!3 \beta^2 $, $\mu_2 = 2\alpha+\beta+2\beta^2 $, $\mu_3= \alpha\!+\!\beta^2$, $\mu_4=3\alpha-3\beta^2/2$, $\mu_5=3(6\alpha-\beta)/4$, $\mu_6=6\alpha+5\beta+9\beta^2$, and $\mu_7=2\alpha-\beta-\beta^2$. At cubic limit with $s_{\theta}=1/\sqrt{3}$ and $c_{\theta}=\sqrt{2/3}$, one obtains $\alpha=\beta=\mu_{1,2,...,7}=0$.

From Eq.~(\ref{eq:p1}), it is evident that $K_{11}$, $\Gamma_{11}$ and $\Gamma'_{11}$ are related to the Hund's coupling and vanish at $J_H=0$ (i.e. $E_1\equiv E_2$), while the Heisenberg $J_{11}$ term remains. In the cubic limit, the exchange parameters are:
\begin{align}
K_{11} &=-\frac{4}{9} \left( \frac{1}{E_1}-\frac{1}{E_2} \right) (3t^2-t'^2),
\notag \\
J_{11} &=\frac{4}{27}\left(\frac{2}{E_2}+\frac{1}{E_3} \right)t'^2 \;,
\notag \\
\Gamma_{11}  &=\frac{8}{9}  \left( \frac{1}{E_1}-\frac{1}{E_2} \right)tt' \;,
\notag \\
\Gamma'_{11}  &= 0 \;,
\label{eq:c1}
\end{align}
which are consistent with previous work \cite{Rau14b}. It is clear that $J_{11}$ and $\Gamma_{11}$ are both positive with the magnitudes related to the direct hopping $t'$. $\Gamma'_{11}=0$ is dictated by cubic symmetry. $K_{11}$ is FM since the indirect hopping $t$ is generally stronger than the direct hopping $t'$ in real materials. $J_{11}$ is AFM and proportional to $t'^2$, while $\Gamma_{11}$ is positive and linear in $t'$.

Once the trigonal crystal field $\Delta$ is introduced, all these four exchange parameters are affected as shown in Fig.~\ref{fig:4}(a). $K_{11}$ is slightly suppressed but remains FM in a wide range of $\delta=2\Delta/\lambda$, and the Heisenberg interaction $J_{11}$ changes from AFM to FM at small negative $\delta$.
The off-diagonal $\Gamma_{11}$ is quite robust and remains positive in the presented range of $\delta$. $\Gamma'_{11}$ term gradually emerges at finite $\delta$ when the orbital degeneracy is lifted.

\subsubsection{2.1.2. Charge-transfer processes}
The virtual excitation of the type $d^5_i d^5_j \rightarrow d^6_i  d^6_j$ is considered for the charge-transfer processes, where two holes are created on the same ligand ion as shown in Fig.~\ref{fig:2}(c). The resulting spin-orbital exchange Hamiltonian along $\gamma=z$-type NN bonds can be written as
\begin{align}
\mathcal{H}^{(z)}_{12}= &-4t^2\left( \frac{P_T}{E_4} +\frac{P_S}{E_5} \right)(n_{ic}+n_{jc})
\notag \\
&-\frac{8 t^2}{3}\left(\frac{2}{E_5}+ \frac{1}{E_6}
\right)P_S (n_{ia}n_{jb}+n_{ib}n_{ja}) \;,
\end{align}
where the excitation energies are $E_4=2\Delta_{pd}+U_P-3J_H^P$,
$E_5=2\Delta_{pd}+U_P-J_H^P$, and $E_6=2\Delta_{pd}+U_P+2J_H^P$. Here
$U_P$ and $J_H^P$ are the Coulomb interaction and Hund's coupling on the
$p$ orbitals of ligand ions.

After projection onto pseudospin doublet Eq. (\ref{eq:wf}), we obtain the following contributions to the exchange parameters:
\begin{align}
K_{12}=& + \frac{16}{9} \left(1+9 \alpha \right) \frac{t^2}{E_{CT}}  \;,
\notag \\
J_{12} =& - \frac{8}{9} (1+6\mu_3 ) \frac{t^2}{E_{CT}}
 +\frac{16}{3}\beta^2\left( \frac{2}{E_5}+\frac{1}{E_6} \right)t^2 \;,
 \notag \\
\Gamma_{12} =& - \frac{4}{3}\mu_8 \frac{t^2}{E_{CT}}
 -\frac{4}{3}\mu_9 \left(\frac{1}{E_4}-\frac{1}{E_5} \right) t^2  \;,
\notag \\
\Gamma'_{12}=&+ \frac{4}{3} \mu_2 \frac{t^2}{E_{CT}}
 -\frac{4}{3}\mu_9  \left(\frac{1}{E_4}-\frac{1}{E_5} \right) t^2  \;,
 \label{eq:p2}
\end{align}
where $\mu_8=\mu_1-\mu_2$, $\mu_9=\mu_2-\mu_7$, and $1/E_{CT}$ is shorthand notation for $1/E_4+  1/3E_5+2/3 E_6 $.

In the cubic limit, the exchange parameters are
\begin{align}
K_{12}=-2J_{12} >0 \;, \ \ \ \ \ \Gamma_{12}=\Gamma'_{12}=0 \;,
\end{align}
the Kitaev interaction $K_{12}$ is AFM and the Heisenberg term $J_{12}$ is FM. Away from the cubic limit, $K_{12}$ ($J_{12}$) remains AFM (FM) in the $\delta$ window shown in Fig.~\ref{fig:4}(b). Both $\Gamma_{12}$ and $\Gamma'_{12}$ are generated by finite $\delta$, and the strength of $\Gamma_{12}$ is rather weak among the four exchange parameters.

\subsubsection{2.1.3. Cyclic exchange}
The same virtual excitation of the type $d^5_i d^5_j \rightarrow d^6_i  d^6_j$ as in charge-transfer processes is considered here. The difference is that two holes are created on different ligand sites and do not meet each other during the cyclic exchange, see Fig.~\ref{fig:2}(d). The obtained spin-orbital exchange Hamiltonian along $\gamma=z$-type NN bonds is
\begin{align}
\mathcal{H}^{(z)}_{13}= \frac{4t^2}{\Delta_{pd}}(\vc S_i \cdot \vc S_j +\tfrac{1}{4})
(a_i^{\dag}b_ia_j^{\dag}b_j+b_i^{\dag}a_ib_j^{\dag}a_j) \;,
\end{align}
which gives the following exchange parameters between pseudospins:
\begin{align}
K_{13}=& -\frac{16}{9}  \left(1+9 \alpha \right) \frac{t^2 }{\Delta_{pd}}  \;,
\notag \\
J_{13} =&\left(\frac{8}{9} + 8 \alpha + \frac{2 \beta }{3 }  \right) \frac{t^2 }{\Delta_{pd}}\;,
\notag \\
\Gamma_{13} =&+\frac{4}{3}\mu_8 \frac{t^2 }{\Delta_{pd}}  \;,
\notag \\
\Gamma'_{13}=&-\frac{4}{3} \mu_2 \frac{ t^2 }{\Delta_{pd}} \;.
 \label{eq:p2}
\end{align}

In the cubic limit, the exchange parameters are
\begin{align}
K_{13}=-2J_{13} <0 \;, \ \ \ \ \ \Gamma_{13}=\Gamma'_{13}=0 \;.
\end{align}
The Kitaev interaction $K_{13}$ is FM, the Heisenberg term $J_{13}$ is AFM, while both off-diagonal terms are zero.

\begin{figure*}
\begin{center}
\includegraphics[width=16cm]{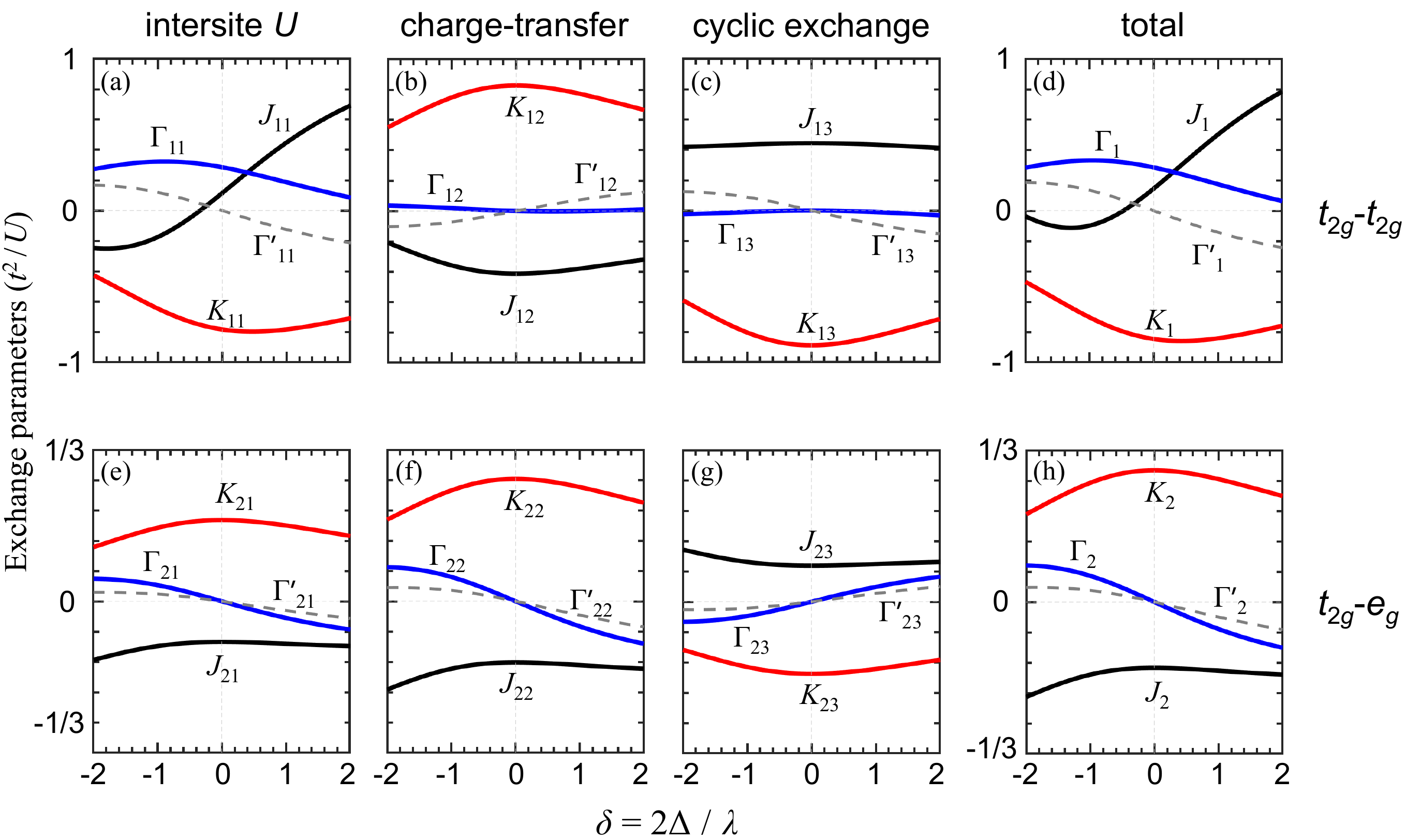}
\caption{Exchange parameters $K$ (red), $J$ (black), $\Gamma$ (blue) and $\Gamma'$ (grey dashed) in units of $t^2/U$ as a function of trigonal field parameter $\delta= 2\Delta/\lambda$ for different hopping processes in $t_{2g}$-$t_{2g}$ (top) and $t_{2g}$-$e_g$ (bottom) channels. Other microscopic parameters are $t'/t=0.5$, $U/\Delta_{pd}=0.5$, $10Dq/U=1$, $J_H/U=0.15$,
$U_p/U=0.6$, $J^p_H/U_p=0.4$ and $t_{pd\sigma}/t_{pd\pi}=2$.}
\label{fig:4}
\end{center}
\end{figure*}

Once the trigonal crystal field is finite, see Fig.~\ref{fig:4}(c), $\Gamma_{13}$ is negligible; $K_{13}$, $J_{13}$, and $\Gamma'_{13}$ have opposite signs of those from charge-transfer processes for the same $\delta$, which lead to cancellations between these couplings.

\subsubsection{2.1.4. Total contributions from hoppings within $t_{2g}$ orbitals}
At this point, we can sum up the contributions to exchange parameters derived from $t_{2g}$-$t_{2g}$ hoppings:
\begin{alignat} {2}
K_1 &= K_{11}+K_{12}+K_{13} \;, & \ \ \ \ J_1 &= J_{11}+J_{12}+J_{13}\;,
\notag \\
\Gamma_1 &= \Gamma_{11}+\Gamma_{12}+\Gamma_{12} \;, &
\Gamma'_1 &= \Gamma'_{11}+\Gamma'_{12}+\Gamma'_{13} \;.
\end{alignat}

As shown in Fig.~\ref{fig:4}(d), the FM $K_1$ interaction is dominant and trigonal field $\Delta$ can tune the Heisenberg interaction $J_1$ from FM to AFM. $\Gamma_1$ coupling remains positive within the presented parameter window and changes sign when approaching larger positive $\delta=2\Delta/\lambda$. The magnitude of $\Gamma'_1$ is proportional to the value of $\delta$. The general trend of the exchange parameters as a function of $\delta$ is similar to that in  Fig.~\ref{fig:4}(a), due to the cancellations between contributions from the charge-transfer and cyclic exchange processes.

\subsection{2.2. Exchange between $t_{2g}$ and empty $e_g$ orbitals}
In the edge-shared octahedra with 90$^{\circ}$ hopping geometry, the overlap between $t_{2g}$ and $e_g$ orbitals is quite large since it involves the $\sigma$-type $pd$-hopping with the amplitude $t_{pd\sigma}$ larger than $t_{pd\pi}$, see Fig.~\ref{fig:3}(d). Therefore it is essential to include the exchange processes between $t_{2g}$ and $e_g$ orbitals. Closely following the above steps, we will present the three relevant processes contributing to the exchange interactions.

\subsubsection{2.2.1 Intersite $U$ processes }
\label{te1}
The intersite $U$ processes between $t_{2g}$ and empty $e_g$ orbitals have been discussed previously \cite{Kha05,Cha13,Foy13}, assuming cubic limit of $\Delta = 0$. The obtained spin-orbital exchange Hamiltonian for $z$-type NN bonds involves  $c= d_{xy}$ orbital and reads as
\begin{align}
\mathcal{H}^{(z)}_{21}=- t_e^2 \left(\frac{P_T}{E_7}
+\frac{P_S}{E_8}\right)(n_{ic}+n_{jc}) \;,
\label{eq:H21}
\end{align}
where $t_e =t\cdot (t_{pd\sigma}/t_{pd\pi})$, and the effective excitation energies are
\begin{align}
E_7 =\left( \frac{\Delta_T}{ \Delta_{pd}} \right)^2 U_T \;,
 \ \ \ \ \ \ \ \
E_8 =\left(  \frac{\Delta_S}{\Delta_{pd} } \right)^2 U_S \;.
\end{align}
Here, we introduced the excitation energies $\Delta_T=\Delta_{pd}+10Dq-2J_H$ and $U_T=U+10Dq-2J_H$ for transitions into virtual triplet $S=1$ states, and $\Delta_S=\Delta_{pd}+10Dq$ and $U_S=U+10Dq$ for transitions into $S=0$ singlet states.

Using the projection tables in Eqs.~(\ref{eq:tai})-(\ref{eq:taf}), one can obtain the corresponding exchange parameters valid at arbitrary $\Delta$ values:
\begin{align}
K_{21}  &= + \frac{4}{9} \left( 1+ 9 \alpha \right)
\left( \frac{1}{E_7}-\frac{1}{E_8} \right)t_e^2 \;,
\notag \\
J_{21}  &=- \frac{2}{9}  \left( 1+6 \mu_3 \right)
 \left( \frac{1}{E_7}-\frac{1}{E_8} \right)t_e^2 \;,
\notag \\
\Gamma_{21} & =-\frac{2}{3} \mu_2
\left( \frac{1}{E_7}-\frac{1}{E_8} \right)t_e^2 \;,
\notag \\
\Gamma'_{21} &=\frac{1}{3} \mu_7
\left( \frac{1}{E_7}-\frac{1}{E_8} \right)t_e^2    \;.
\label{eq:pe1}
\end{align}
The exchange couplings in Eq.~(\ref{eq:pe1}) are due to Hund's coupling and all vanish when $J_H=0$ (i.e. $E_7\equiv E_8$). For finite $J_H$, we have the following relations at cubic limit:
\begin{align}
K_{21}=-2J_{21} >0 \;, \ \ \ \ \ \Gamma_{21}=\Gamma'_{21}=0 \;.
\end{align}

Under trigonal crystal field, $K_{21}$ ($J_{21}$) remains AFM (FM), while the off-diagonal terms $\Gamma_{21}$ and $\Gamma'_{21}$ are relatively weak, see Fig.~\ref{fig:4}(e). The overall magnitudes of the exchange parameters are smaller compared with the contributions from hoppings between $t_{2g}$ and $t_{2g}$ orbitals, because of the larger excitation energies when the empty $e_g$ orbitals are involved.

\subsubsection{2.2.2. Charge-transfer processes}
Considering the charge-transfer processes between $t_{2g}$ and $e_g$ orbitals, one can obtain the spin-orbital exchange Hamiltonian for $z$-type NN bonds:
\begin{align}
\mathcal{H}^{(z)}_{22}=- t_e^2\left(\frac{P_T}{E_9}
+\frac{P_S}{E_{10}}\right)(n_{ic}+n_{jc}) \;,
\label{eq:H22}
\end{align}
where the effective excitation energies are
\begin{align}
E_9 &=\frac{1}{2}\left( \frac{\Delta_T}{\Delta_{pd}+\Delta_T}  \right)^2 (\Delta_{pd}+\Delta_T+U_p) \;,
\notag \\
E_{10} &=\frac{1}{2}\left( \frac{\Delta_S}{\Delta_{pd}+\Delta_S} \right)^2 (\Delta_{pd}+\Delta_S+U_p) \;.
\end{align}
After projection of Hamiltonian Eq.~(\ref{eq:H22}) onto pseudospin doublet, one obtains the exchange parameters as follows:
\begin{align}
K_{22}  &= + \frac{4}{9} \left( 1+ 9 \alpha \right)
\left( \frac{1}{E_9}-\frac{1}{E_{10} } \right)t_e^2 \;,
\notag \\
J_{22}  &=- \frac{2}{9}  \left( 1+6 \mu_3 \right)
 \left( \frac{1}{E_9}-\frac{1}{E_{10} } \right)t_e^2 \;,
\notag \\
\Gamma_{22} & =-\frac{2}{3} \mu_2
\left( \frac{1}{E_9}-\frac{1}{E_{10} } \right)t_e^2 \;,
\notag \\
\Gamma'_{22} &=\frac{1}{3} \mu_7
\left( \frac{1}{E_9}-\frac{1}{E_{10} } \right)t_e^2    \;.
\label{eq:pe2}
\end{align}

This contribution again is Hund's coupling effect as in Sec. 2.2.1. In the cubic limit, we have
\begin{align}
K_{22}=-2J_{22} >0 \;, \ \ \ \ \ \Gamma_{22}=\Gamma'_{22}=0 \;.
\end{align}
AFM $K_{22}$ and FM $J_{22}$ are very robust against trigonal field parameter $\delta$, and the relatively weak off-diagonal terms $\Gamma_{22}$ and $\Gamma'_{22}$ are generated at the same time, see Fig.~\ref{fig:4}(f).

\subsubsection{2.2.3. Cyclic exchange}
The spin-orbital exchange Hamiltonian of cyclic exchange processes between $t_{2g}$ and $e_g$ orbitals
for $z$-type NN bonds is:
\begin{align}
\mathcal{H}^{(z)}_{23}= t_e^2 \left(\frac{P_T}{E_{11}}
+\frac{P_S}{E_{12}}\right)(n_{ic}+n_{jc}) \;.
\label{eq:H23}
\end{align}
The only active orbitals are $c=d_{xy}$ ones again. The effective excitation energies are:
\begin{align}
E_{11} &=\left(  \frac{\Delta_T}{\Delta_{pd}+\Delta_T}  \right)^2  \Delta_T  \;,
\notag \\
E_{12} &=\left( \frac{\Delta_S}{\Delta_{pd}+\Delta_S}  \right)^2 \Delta_S \;.
\end{align}
The opposite overall sign in Eq.~(\ref{eq:H23}) compared with Eq.~(\ref{eq:H21}) and Eq.~(\ref{eq:H22}) originates from the overlap phase factor between $p$ and $e_g$ orbitals, as illustrated in Fig.~\ref{fig:3}(d).

The obtained exchange parameters are:
\begin{align}
K_{23}  &=- \frac{4}{9} \left( 1+ 9 \alpha \right)
\left( \frac{1}{E_{11}}-\frac{1}{E_{12} } \right)t_e^2 \;,
\notag \\
J_{23}  &=+ \frac{2}{9}  \left( 1+6 \mu_3 \right)
 \left( \frac{1}{E_{11}}-\frac{1}{E_{12} } \right)t_e^2 \;,
\notag \\
\Gamma_{23} & =+\frac{2}{3} \mu_2
\left( \frac{1}{E_{11}}-\frac{1}{E_{12} } \right)t_e^2 \;,
\notag \\
\Gamma'_{23} &=-\frac{1}{3} \mu_7
\left( \frac{1}{E_{11}}-\frac{1}{E_{12} } \right)t_e^2    \;.
\label{eq:pe3}
\end{align}
All four parameters have the opposite signs to those from the intersite $U$ processes Eq.~(\ref{eq:pe1}) between $t_{2g}$ and $e_g$ orbitals, see Figs.~\ref{fig:4}(e) and \ref{fig:4}(g).

\subsubsection{2.2.4. Total contributions from hoppings between $t_{2g}$ and $e_g$ orbitals}
Summing up all the contributions involving $e_g$ orbitals, one finds:
\begin{alignat} {2}
K_2 &= K_{21}+K_{22}+K_{23} \;, & \ \ \ \ J_2 &= J_{21}+J_{22}+J_{23}\;,
\notag \\
\Gamma_2 &= \Gamma_{21}+\Gamma_{22}+\Gamma_{22} \;, &
\Gamma'_2 &= \Gamma'_{21}+\Gamma'_{22}+\Gamma'_{23} \;.
\end{alignat}

As shown in Fig.~\ref{fig:4}(h), the resulting $K_2$ is AFM and $J_2$ is FM, both $\Gamma$ and $\Gamma'$ change from positive to negative when the trigonal field parameter $\delta$ changes from negative to positive values. Due to the cancellation between intersite $U$ and cyclic exchange processes, the total contribution of exchange parameters in Fig.~\ref{fig:4}(h) is very similar to that from charge-transfer contribution in Fig.~\ref{fig:4}(f).

\subsection{2.3. Exchange parameters}

\begin{figure}
\begin{center}
\includegraphics[width=8.5cm]{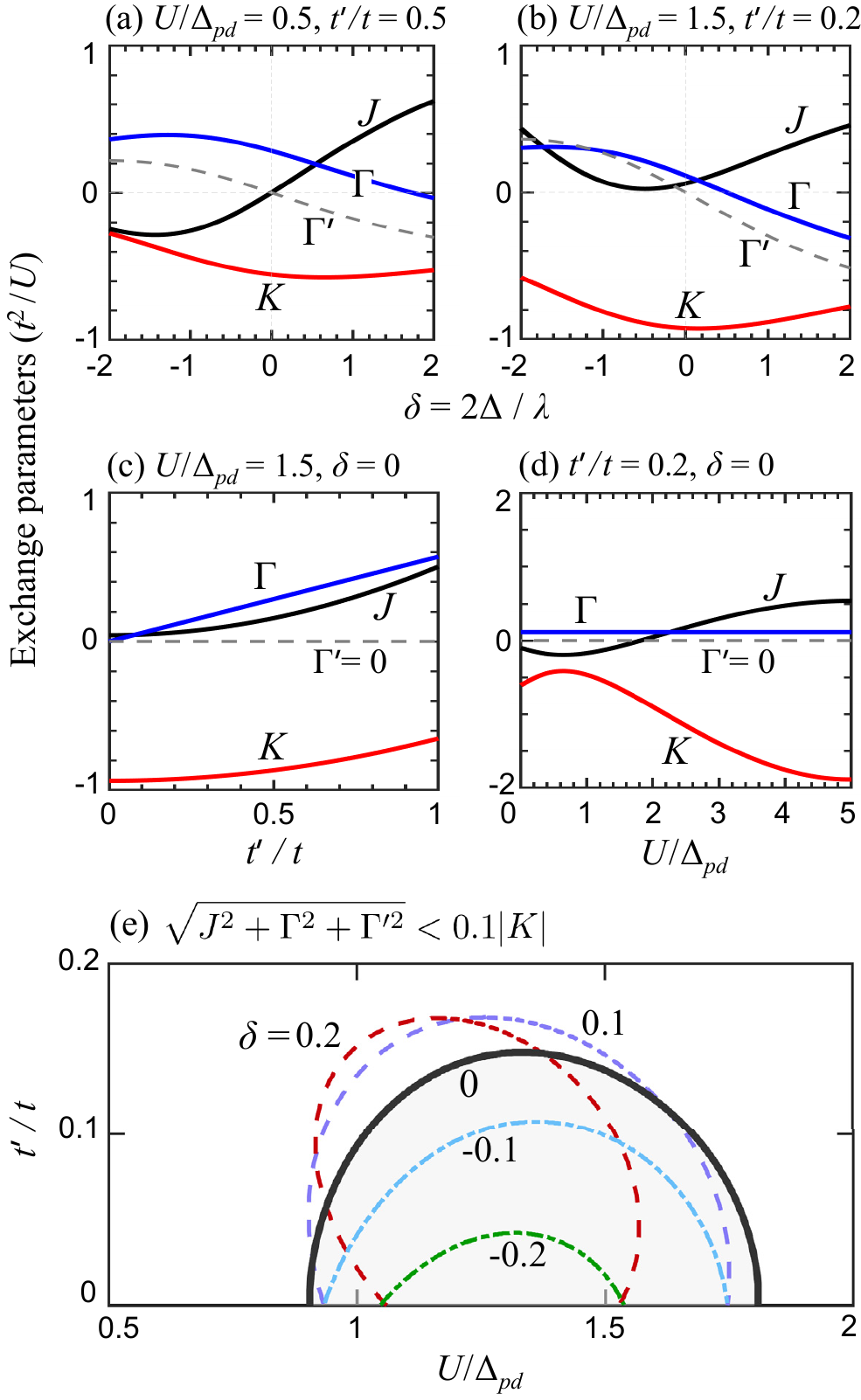}
\caption{Exchange parameters $K$ (red), $J$ (black), $\Gamma$ (blue), and $\Gamma'$ (grey dashed) as a function of (a)-(b) $\delta= 2\Delta/\lambda$, (c) $t'/t$, and (d) $U/\Delta_{pd}$. Other microscopic parameters $10Dq/U=1$, $J_H/U=0.15$, $U_p/U=0.6$, $J^p_H/U_p=0.4$, and $t_{pd\sigma}/t_{pd\pi}=2$ are fixed. (e) The framed areas indicate the parameter subspace of $t'/t$ and $U/\Delta_{pd}$, where the non-Kitaev interaction terms are by an order of value smaller than the FM Kitaev coupling, i.e. $\sqrt{J^2+\Gamma^2+\Gamma'^2}<0.1|K|$. The enclosed area of dominant Kitaev interaction varies as function of $\delta$ and is largest near the cubic limit. }
\label{fig:5}
\end{center}
\end{figure}

Having quantified all the essential exchange channels, we can write the final exchange constants as
\begin{alignat} {2}
K  &= K_1+K_2 \;,
& \ \ \ \ \ \ \ \
J &= J_1+J_2 \;,
\notag \\
\Gamma &= \Gamma_1+\Gamma_2 \;,
&
\Gamma' &=\Gamma'_1+\Gamma'_2    \;.
\label{eq:par}
\end{alignat}

We sum up the results in Fig.~\ref{fig:4}(d) and Fig.~\ref{fig:4}(h), and present the resulting total values of the exchange parameters in Fig.~\ref{fig:5}(a). The following general features can be observed here: \\
\textbf{(i)} the Kitaev interaction $K$ remains FM and is dominant at small $\delta$ regime;
\vskip 2mm
\noindent
\textbf{(ii)} the Heisenberg interaction $J$ can be manipulated between AFM and FM by the trigonal field parameter $\delta$, and becomes comparable with $K$ at large $|\delta|$ regime;
\vskip 2mm
\noindent
\textbf{(iii)} positive $\Gamma$ term changes to negative values only for large positive $\delta$ values; and
\vskip 2mm
\noindent
\textbf{(iv)} $\Gamma'$ term is generated by finite trigonal crystal field and changes the sign when reversing $\delta$.

We notice that the general trend of the exchange couplings in Fig.~\ref{fig:5}(a) is similar with those in Fig.~\ref{fig:4}(d) from exchange between $t_{2g}$ and $t_{2g}$ orbitals. This is due to the larger overall magnitude of the exchange parameters generated within the $t_{2g}$-$t_{2g}$ channel when compared to those of the $t_{2g}$-$e_g$ channel. Thus the major features of the $t_{2g}$-$t_{2g}$ contributions are preserved. However, the contributions involving $e_g$ orbitals modify the quantitative values of the exchange parameters, which is very important especially when determining the proximity of a given compound to the Kitaev spin liquid phase.

Besides the trigonal crystal field, the exchange parameters also depend on several other microscopic parameters such as $J_H/U$, $t'/t$, and $U/\Delta_{pd}$. For transition metal ions, Hund's coupling values of the order of $J_H/U \sim 0.1-0.2$ are typical \cite{Suz21,Foy13,Kim14,Pra59,Ani91,Pic98,Jia10}. However, $t'/t$ and $U/\Delta_{pd}$ values may vary broadly among transition metal compounds. For comparison, we show the exchange parameters with smaller $t'/t$ and larger $U/\Delta_{pd}$ values in Fig.~\ref{fig:5}(b); the FM $K$ is greatly enhanced than that in Fig.~\ref{fig:5}(a), which can increase the possibility of realizing the spin liquid phase.

To illustrate the effects of $t'/t$ and $U/\Delta_{pd}$ in more detail, we here present the exchange constants as a function of them in Figs.~\ref{fig:5}(c) and ~\ref{fig:5}(d), respectively. The cubic limit $\delta=0$ resulting in $\Gamma'=0$ is shown as a representative. Also, one can get $\Gamma=\Gamma_{11}=\tfrac{8}{9}  \left( \tfrac{1}{E_1}-\tfrac{1}{E_2} \right)tt'$  from Eq.~(\ref{eq:c1}) in the cubic limit, which indicates $\Gamma$ is linear in $t'/t$ and independent of $U/ \Delta_{pd}$. Figure~\ref{fig:5}(c) shows that the direct $t'$-hopping contributions to $K$ and $J$ are proportional to $t'^2$. This can also be inferred from Eq.~(\ref{eq:c1}) which gives $K_{11} \simeq - 8J_H(3t^2-t'^2)/9U^2$ and $J_{11} \simeq 4t'^2/9U$.It is important to observe in Fig.~\ref{fig:5}(d) that FM K is strongly enhanced at charge-transfer limit ($U \gg \Delta_{pd}$), and the Heisenberg $J$ can be switched from FM to AFM when increasing $U/ \Delta_{pd}$.

Our results suggest that materials with small $\delta$ and $t'/t$ ratio, hence small $J$, $\Gamma$, and $\Gamma'$ terms, provide more favorable conditions for realization of the Kitaev model. To show this, we plot in Fig.~\ref{fig:5}(e) the parameter space with dominant Kitaev interaction; more specifically, we show the areas where the non-Kitaev terms are less than 10\% of Kitaev coupling: $\sqrt{J^2+\Gamma^2+\Gamma'^2}<0.1|K|$. This plot suggests that materials with nearly cubic symmetry (small $\delta$), and close to the charge-transfer limit ($U/\Delta_{pd}\sim$1-2) such as Co$^{4+}$ systems, would be promising candidates to realize the Kitaev spin liquid.

\section{III. Implications for $\text{Na}_2\text{IrO}_3$ and $\text{RuCl}_3$ }
\textcolor[rgb]{0.00,0.00,0.00}{Having quantified all the exchange contributions as functions of various microscopic parameters, we now apply our theory to two representative Kitaev materials, Na$_2$IrO$_3$ and RuCl$_3$, which have extensively been studied in recent years. While the exchange parameters obtained from different experimental data sets and their model fits vary quite significantly (see, e.g., Refs.~\cite{Lau20,Mak20}), they generally agree on the following points common to both compounds: (i) the largest term is given by FM Kitaev coupling $K<0$, (ii) the next leading terms are $\Gamma$ and $J$ couplings, and (iii) $\Gamma'$ and longer-range (e.g., third-NN Heisenberg $J_3$) couplings are much smaller than the $K$ term but need to be included in the detailed data fits. Quantitatively, however, there are some essential differences between the exchange parameters in Na$_2$IrO$_3$ and RuCl$_3$, with important implications for their physical properties. In particular, while the zigzag-type magnetic correlations are very robust in Na$_2$IrO$_3$, persisting far above N\'{e}el temperature $T_N$ \cite{Chu15,Kim20}, they are very fragile in RuCl$_3$ and get readily destabilized above $T_N$ by competing FM-type correlations~\cite{Suz21}. The physical origin of these contrasting features is discussed below.}

\begin{figure}
\begin{center}
\includegraphics[width=7cm]{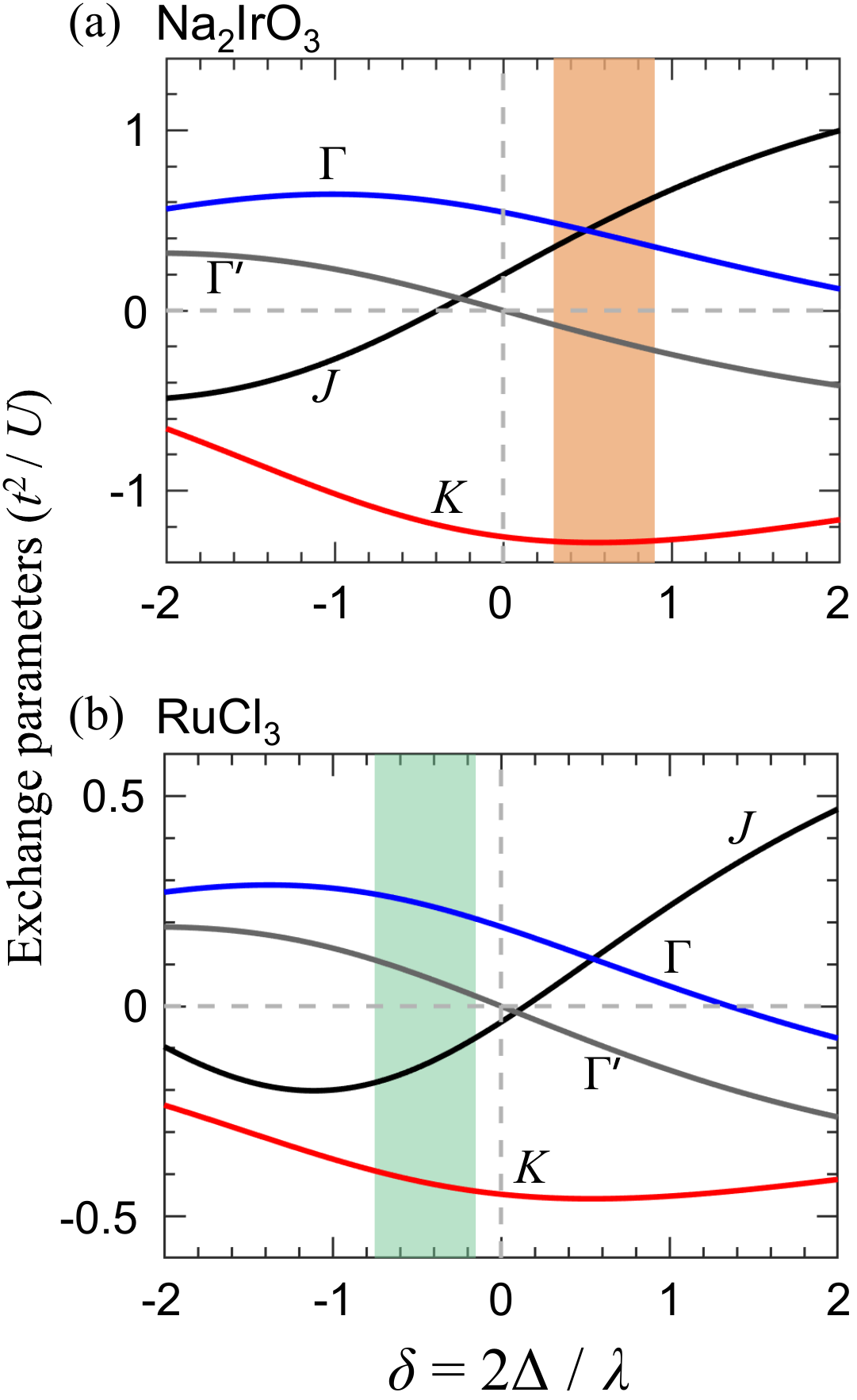}
\caption{Exchange parameters $K$ (red), $J$ (black), $\Gamma$ (blue), and $\Gamma'$ (grey) as a function of $\delta= 2\Delta/\lambda$, calculated for (a) Na$_2$IrO$_3$ and (b) RuCl$_3$. For Na$_2$IrO$_3$, we use $10Dq=3.3$ eV, $U=1.35$ eV, $J_H=0.25$ eV, and $\Delta_{pd}=3.3$ eV values from Ref.~\cite{Kim14}. $U_p=4$ eV and Hund's $J_p=1.6$ eV for O-$p$ orbitals are taken from Ref.~\cite{Foy13}, and we set $t'=0.6t$. For  RuCl$_3$, we use $10Dq=2.4$ eV, $J_H=0.34$ eV, $\Delta_{pd}=5.5$ eV and $U=2.5$ eV \cite{Suz21}. $U_p=1.5$ eV and Hund's $J_p=0.7$ eV are used for Cl-$p$ orbitals, and $t'=0.5t$. The shaded intervals cover the $\delta$ values suggested by the experimental data on Na$_2$IrO$_3$ \cite{Sin10,Gre13} and RuCl$_3$ \cite{Agr17,Suz21}.}
\label{fig:6}
\end{center}
\end{figure}

\subsection{3.1 $\text{Na}_2\text{IrO}_3$}
Figure~\ref{fig:6}(a) presents the calculated exchange parameters as a function of trigonal crystal field parameter $\delta=2\Delta/\lambda$, using the microscopic parameters suitable for Na$_2$IrO$_3$. The results show that $\delta$ is a very important parameter to control the exchange couplings. The value of this parameter can be inferred from the $g$-factor anisotropy, or the splitting of spin-orbit exciton levels. In Na$_2$IrO$_3$, a rather wide interval of $\delta \sim$ 0.3-0.9 have been suggested~\cite{Sin10,Gre13}.

Considering a representative value of $\delta=0.75$,
\textcolor[rgb]{0.00,0.00,0.00}{
we obtain ($K,J,\Gamma,\Gamma'$) $\simeq$ ($-1.3, 0.57, 0.39, -0.2$) $t^2/U$ for Na$_2$IrO$_3$ from our theory. The NN exchange Hamiltonian is dominated by FM Kitaev term $K \simeq -1.3 \; t^2/U$, followed by AF Heisenberg $J/|K|\sim 0.4$ and positive $\Gamma/|K|\sim 0.3$ couplings, and a small negative $\Gamma'/|K|\sim -0.15$ term. }

\textcolor[rgb]{0.00,0.00,0.00}{
Recently, the following exchange parameters have been reported for Na$_2$IrO$_3$, based on the analysis of the resonant inelastic x-ray scattering (RIXS) data~\cite{Kim20}: ($K,J,\Gamma,\Gamma'$) $\simeq$ ($-24, 12, 11, -3$) meV (see the $A2$ parameter set in Ref.~\cite{Kim20}). To compare our theory with these experimental values, we set overall energy scale of $t^2/U \simeq 19$ meV to obtain ($K,J,\Gamma,\Gamma'$) $\simeq$ ($-24, 10.6, 7.2, -3.6$) meV. The signs and relative values of the above exchange couplings are well reproduced. }

\begin{figure}
\begin{center}
\includegraphics[width=8.5cm]{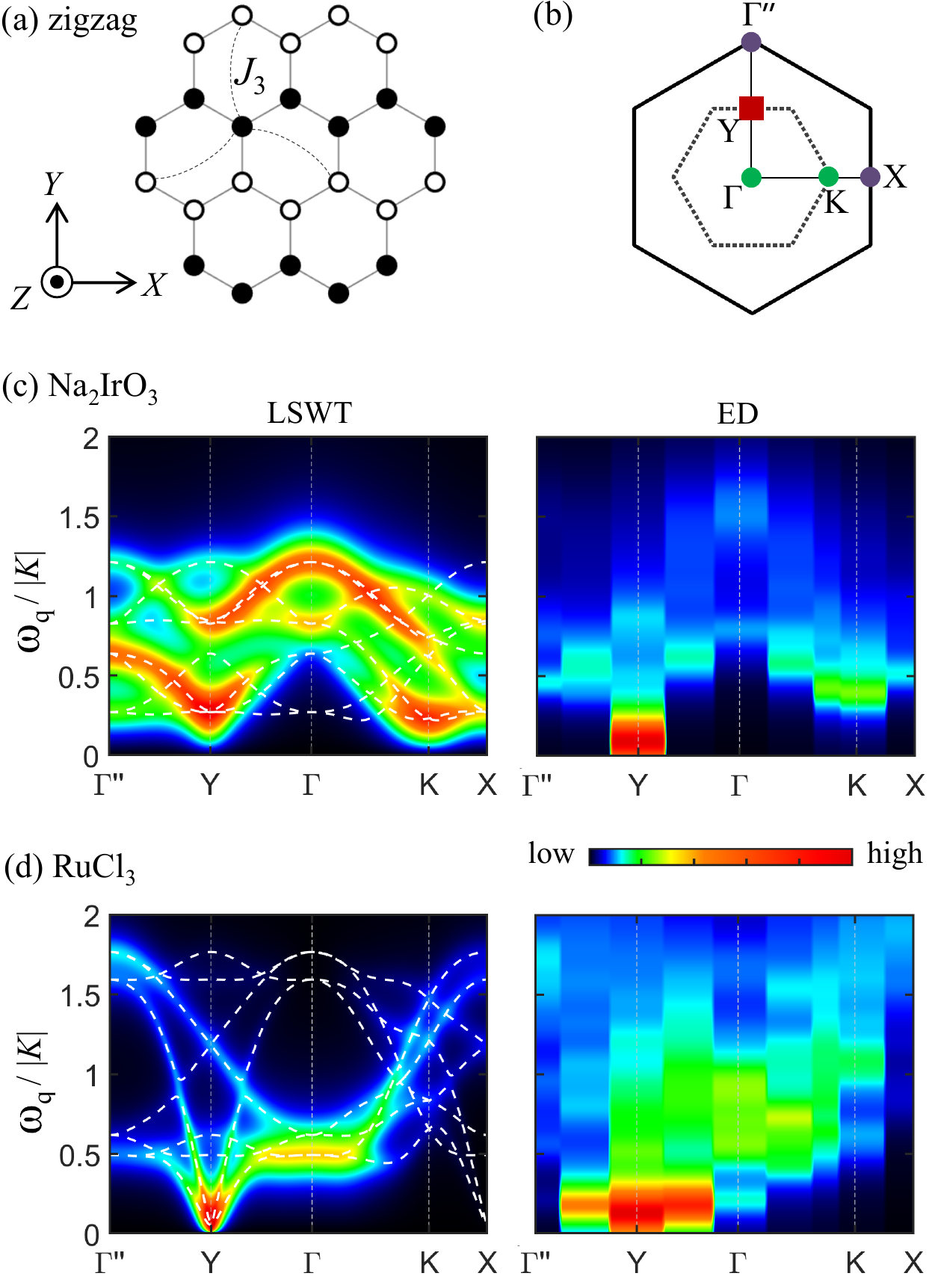}
\caption{(a) Sketch of the magnetic structure for zigzag order. Open and closed circles represent spins with opposite directions. (b) Brillouin zones of the honeycomb (inner hexagon) and the completed triangular lattice (outer hexagon). Expected spin excitation spectrum using linear spin wave theory (LSWT) and the exact diagonalization (ED) with (c) ($K,J,\Gamma,\Gamma',J_3$)=($-1,0.44,0.3,-0.15,0.1$)$|K|$ for Na$_2$IrO$_3$ and (d) ($K,J,\Gamma,\Gamma',J_3$)= ($-1,-0.31,0.56,0.15,0.22$) $|K|$ for RuCl$_3$, along the $\Gamma''$-Y-$\Gamma$-K-X path of the Brillouin zone. Plotted in (c,d) is the trace of the spin susceptibility tensor. The LSWT spectra are averaged over three possible zigzag pattern directions. The ED data are a combination of the results for hexagonal 24-site and 32-site clusters. }
\label{fig:7}
\end{center}
\end{figure}

The Ir moments in Na$_2$IrO$_3$ undergo a zigzag order shown in Fig.~\ref{fig:7}(a) at $T_N \simeq 15$ K \cite{Liu11,Cho12,Ye12}.
It has been pointed out a while ago~\cite{Cha16} that the orientation of the ordered moments imposes a strong constraint on the possible signs of the $K$ and $\Gamma$ couplings. In Na$_2$IrO$_3$, and as a matter of fact also in RuCl$_3$, the moments were observed~\cite{Chu15,Sea20} to be confined to the $XZ$ plane (i.e. crystallographic $ac$ plane) and pointing between two ligand (O or Cl) ions. This observation dictates the sign combination of $K<0$ and $\Gamma>0$ (see Fig.~1 of Ref.~\cite{Cha16}). More quantitatively, the angle $\alpha$ between the ordered moment direction and $X\parallel a$ axis is, on a classical level, given by
\begin{align}
\tan 2\alpha=\frac{4\sqrt{2}(-K+\Gamma-\Gamma')}{2K+7\Gamma+2\Gamma'} \;,
\label{eq:an}
\end{align}
which depends also on (small) $\Gamma'$ parameter. Using the theoretical values of $K$, $\Gamma$, and $\Gamma'$ calculated above, we estimate $\alpha=43^{\circ}$ in Na$_2$IrO$_3$, consistent with the experiment~\cite{Chu15}.

Next, we would like to present the spin excitation spectra calculated using our exchange parameters. At this point, we need to supplement the model with a small third-NN Heisenberg coupling $J_3$ which is known to stabilize zigzag order over the other competing states~\cite{Kim11,Rus19}. It is difficult to estimate $J_3$ value analytically, since the long-range interactions involve multiple exchange channels. Instead, they are more reliably obtained from experimental fits. Here, we adopt a value of $J_3=0.1|K|=2.4$ meV for Na$_2$IrO$_3$, similar to that used in Ref.~\cite{Kim20} when fitting the experimental RIXS data.

The expected spin excitation spectra are calculated using the linear spin wave theory (LSWT) and exact diagonalization (ED) method. For both LSWT and ED results shown in Fig.~\ref{fig:7}(c), the energy minimum is at Bragg point $q={\rm Y}$. There are no low-energy excitations around $q=\Gamma$ point as in the case of RuCl$_3$ which will be discussed below. Instead, a soft mode near $q={\rm K}$ point is observed, suggesting a competing phase with the characteristic wave vector $q \sim {\rm K}$.
\textcolor[rgb]{0.00,0.00,0.00}{
These findings are in a broad consistency with the spin excitation spectra measured by RIXS~\cite{Kim20}. For a detailed comparison with experiment, however, the neutron scattering data with a higher resolution (than in the current RIXS data) is desirable.}
It should be also noticed that while LSWT and ED results are qualitatively similar, LSWT does not capture the decay processes that lead to large broadening of the magnon spectral functions \cite{Zhi13,Du15,Du16,Win17b}, especially at high energies.

\subsection{3.2 $\text{Ru}\text{Cl}_3$}

The exchange constants as a function of trigonal crystal field splitting $\delta=2\Delta/\lambda$, calculated for microscopic parameters appropriate for RuCl$_3$, are shown in Fig.~\ref{fig:6}(b). These dependencies of the exchange parameters are qualitatively similar to those in Fig.~\ref{fig:6}(a), such that the FM $K$ is dominant at small $\delta$ regime, Heisenberg $J$ changes from FM to AFM when $\delta$ varies, $\Gamma$ remains positive in most of the parameter space, and $\Gamma'$ emerges when the cubic symmetry is broken.

Opposite to Na$_2$IrO$_3$ case with $\delta>0$, negative values of $\delta$ in the range from $\delta \simeq -0.15$ \cite{Agr17} to $ -0.75 $ \cite{Suz21} have been suggested by experiments in RuCl$_3$. Considering  a representative value of $\delta=-0.4$, we find ($K,J,\Gamma,\Gamma'$) $\simeq$ ($-0.42,-0.13,0.24,0.06$) $t^2/U$. The values of non-Kitaev terms (relative to $K$) are sizable, as in case of Na$_2$IrO$_3$. On the other hand, the signs of Heisenberg $J$ and $\Gamma'$ terms of are opposite to those in Na$_2$IrO$_3$. This is due to the opposite signs of crystal field $\delta$ in the two materials, cf. Figs.~\ref{fig:6}(a) and \ref{fig:6}(b). \textcolor[rgb]{0.00,0.00,0.00}{Figure~\ref{fig:6} also suggests that by tuning the trigonal field $\delta$ towards the cubic limit (e.g., by means of the $c$ axis strain), one can nearly suppress $J$ and $\Gamma'$ terms in both compounds, realizing thereby the $K$-$\Gamma$ model of large current interest~\cite{Gor19,Wan19,Bue21}. }

\textcolor[rgb]{0.00,0.00,0.00}{A number of various parameter sets for RuCl$_3$ have been suggested in the literature. As said above, they mostly agree on $K<0$ and $\Gamma>0$ sign combination, which is conclusively evidenced by experiments~\cite{Sea20,Koi20}. One widely used parameter set, inspired by \emph{ab-initio} studies and supported by inelastic neutron scattering data, reads as follows: ($K,J,\Gamma,\Gamma'$) $\simeq$ ($-5, -0.5, 2.5, 0$) meV~\cite{Win18,Win17b}. Recently, these values have been updated to ($K,J,\Gamma,\Gamma'$) $\simeq$ ($-5, -3, 2.5, 0.1$) meV, in order to be consistent with the RIXS data~\cite{Suz21} revealing a close proximity of FM state in RuCl$_3$. To compare our theory with these values, we set the overall energy scale $t^2/U \simeq 12$ meV and obtain ($K,J,\Gamma,\Gamma'$) $\simeq$ ($-5, -1.6, 2.8, 0.7$) meV. This reproduces the signs and overall hierarchy of the measured exchange parameters. We note that a difference in $t^2/U$ scales in Na$_2$IrO$_3$ and RuCl$_3$ (19 vs 12 meV) can be attributed to more extended nature of $5d$ wavefunctions, hence the larger $t$ and smaller $U$ values in iridates. }

Using Eq.~(\ref{eq:an}) with the calculated $K$, $\Gamma$, and $\Gamma'$ values, the ordered moment direction in the zigzag phase is evaluated as $\alpha \simeq 37^{\circ}$ in RuCl$_3$, consistent with the experimental observations~\cite{Sea20,Cao16}.

The zigzag order has been observed in RuCl$_3$~\cite{Sea15,Joh15} below $T_N \simeq 7$ K. Our obtained NN exchange parameters suggest classical FM ground state instead; \textcolor[rgb]{0.00,0.00,0.00}{this is due to FM $J$ as well as positive $\Gamma'$ couplings. We therefore add a small third-NN Heisenberg coupling $J_3=1.1$ meV, which is just enough to stabilize the zigzag order, and calculate the spin excitation spectra. } The results are shown in Fig.~\ref{fig:7}(d). The LSWT result is similar to the data obtained by ED method: a very small magnon gap is opened at the Bragg point $q={\rm Y}$, and the strongest intensity concentrates at the same $q$ point. In contrast to Na$_2$IrO$_3$, the soft mode occurs here near the $\Gamma$ point. This is because of that the $q \sim 0$ correlations, which are typical for the FM Kitaev model, are further enhanced by FM Heisenberg interaction $J<0$ in RuCl$_3$. The presence of low-energy and small $q\sim0$ correlations are consistent with the neutron scattering experiments \cite{Ban16,Ban17,Ban18}. The intensity at $q={\rm Y}$ point will transfer to $q=\Gamma$ point at finite temperatures \cite{Ban17,Do17,Ban18,Suz21}.

These observations imply that magnetic states with characteristic vector $q \sim 0$, such as the ferromagnetic one, are closely competing with the zigzag order in RuCl$_3$~\cite{Suz21}. The proximity to the FM state should be essential to understand the field dependent behavior, especially given the nontrivial topology of ferromagnetic magnons in Kitaev materials~\cite{McC18,Jos18,Che21}.

Classically, the energy difference $\delta E=E_{\rm FM}-E_{\rm ZZ}$ between the FM and zigzag states is:
\begin{equation}
\delta E=\frac{1}{8}\left[ \sqrt{ \Lambda ^2 \! + \! 2 \Gamma^2} + \Lambda +2 (J + 3J_3-2\Gamma') \right],
\label{eq:ediff}
\end{equation}
where $\Lambda=K - \frac{\Gamma}{2}  + \Gamma'$. One can get $\delta E\simeq 0.2$ meV for RuCl$_3$ using our exchange parameter set ($K,J,\Gamma,\Gamma', J_3$) $\simeq$ ($-5, -1.6, 2.8, 0.7, 1.1$) meV. The small value of energy difference $\delta E$ implies the presence of low-energy competing $q \sim 0$ states, and explains the quick suppression of the zigzag order by a magnetic field or temperature in RuCl$_3$.

It is instructive to evaluate an equivalent energy difference for the case of Na$_2$IrO$_3$. Using the exchange couplings \textcolor[rgb]{0.00,0.00,0.00}{($K,J,\Gamma,\Gamma', J_3$) $\simeq$ ($-24, 10.6, 7.2, -3.6, 2.4$) meV considered above, we obtain} $\delta E \simeq 6.4$ meV which is much larger than that in RuCl$_3$. This large difference is due to the strong AFM $J>0$ and negative $\Gamma'<0$ couplings in Na$_2$IrO$_3$, as can be inferred from Eq.~(\ref{eq:ediff}). The large $\delta E$ value implies that there is no close competition between FM and zigzag states, and explains the robustness of zigzag order in Na$_2$IrO$_3$ against magnetic field~\cite{Das19} or temperature~\cite{Chu15,Kim20}.

The above considerations show that the differences between the exchange constants in Na$_2$IrO$_3$ and RuCl$_3$, and thus their different magnetic behaviors, are due to the opposite signs of the non-cubic trigonal crystal field $\Delta$ in these compounds. This illustrates a decisive role of lattice distortion in determining the magnetic properties of Kitaev materials.

\section{V. Conclusions}
In summary, we have derived the exchange Hamiltonian general for $d^5$ spin-orbit Mott insulators with 90$^{\circ}$ bonding exchange geometry. The trigonal crystal field and the hopping channels involving $e_g$ orbitals are included. Generally, we have found that the exchange Hamiltonian is characterized by dominant FM Kitaev and sizable non-Kitaev terms at a very wide range of trigonal crystal field. Our results suggest that a parameter region close to the charge-transfer insulator regime and cubic limit is favored to realize the Kitaev spin liquid phase.

We have applied our theory to two representative Kitaev candidate materials: Na$_2$IrO$_3$ and RuCl$_3$. Both materials were found to have dominant FM $K$ and positive $\Gamma$ terms. The relative strengths of non-Kitaev couplings to the Kitaev term are very similar in these two materials. However, Na$_2$IrO$_3$ has AFM Heisenberg $J$ and negative $\Gamma'$ couplings ($J>0,\Gamma'<0$) while RuCl$_3$ possesses FM $J$ and positive $\Gamma'$ terms ($J<0,\Gamma'>0$), \textcolor[rgb]{0.00,0.00,0.00}{with important implications for their physical properties such as the stability of the zigzag order and magnetic excitation spectra.} Our calculations reveal that the qualitative differences of $J$ and $\Gamma'$ exchange constants between Na$_2$IrO$_3$ and RuCl$_3$ originate from the opposite signs of the trigonal crystal field in these two compounds. This suggests that the magnetic properties of Kitaev materials can efficiently be manipulated by tuning the crystal field, e.g., via strain or pressure control.

\section{Acknowledgments}

H. L. thanks Z. Z. Du for useful discussions, and acknowledges support by the W$\ddot{\rm u}$rzburg-Dresden Cluster of Excellence on Complexity and Topology in Quantum Matter — \emph{ct.qmat} (EXC 2147, project ID  390858490). H. L. and G. Kh. acknowledge support by the European Research Council under Advanced Grant No. 669550 (Com4Com). J. Ch. acknowledges support by Czech Science Foundation (GA\v{C}R) under Project No.~GA19-16937S. Computational resources were supplied by the project ``e-Infrastruktura CZ'' (e-INFRA LM2018140) provided within the program Projects of Large Research, Development and Innovations Infrastructures.

\end{document}